\documentclass[letterpaper, 10 pt, conference]{ieeeconf}

\IEEEoverridecommandlockouts                              
\pdfoutput=1 

\usepackage{graphicx,subfig,cite,ifpdf}
\usepackage{amsmath,amssymb,array}
\usepackage{stmaryrd,algorithmic,algorithm}
\usepackage{enumerate}
\usepackage[singlelinecheck=false]{caption}

\newtheorem{remark}{Remark}

\title{\LARGE \bf
Encrypted Control System with Quantizer}

\author{Masako Kishida
\thanks{M. Kishida is with National Institute of Informatics / address: 2-1-2 Hitotsubashi, Chiyoda-ku, Tokyo 101-8430, Japan  / phone: +81-3-4212-2231  / email: {\tt\small kishida@nii.ac.jp}}
}

\begin{document}

\maketitle
\thispagestyle{empty}
\pagestyle{empty}

\begin{abstract} 
This paper considers the design of encrypted control systems to secure data privacy when the control systems operate over a network.
In particular, we propose to combine Paillier cryptosystem with a quantizer whose sensitivity changes with the evolution of the system.
This allows the encrypted control system to balance between the cipher strength and processing time.
Such an ability is essential for control systems that are expected to run real-time.  
It also allows the closed-loop system to achieve the asymptotic stability for linear systems. 
Extensions to event-triggered control and nonlinear control systems are also discussed.
\end{abstract}

\section{Introduction}

Networked control systems are ubiquitous \cite{SanAJ15,ZhaGK13,BulCM09,AntB07}.
The use of networks has not only reduced the deployment cost and increased the flexibility of control systems, but also allowed the systems to 
outsource computations of control inputs to a cloud controller when a plant does not have sufficient computational resources \cite{TakSS17,FarSB16}. 
However, this raises new privacy and security concerns; plants may want to protect the privacy of the sampled data because the cloud controller is not trustworthy, and communication networks may be vulnerable to cyber-attacks \cite{SanAJ15}.

One approach to protect privacy is to use ``differential privacy''  \cite{Dwo06}.
Differential privacy adds noise to the data so that 
 the contribution of a specific agent is hidden without changing the solution to the problem significantly.
Although differential privacy is a relatively new notion, it has found applications in a variety of networked systems including systems and controls \cite{HuaD12,CorDH16,MoM17}.

Another approach is to use ``encryptions''. ``Encrypted control system'' is a control architecture in which the controller computes the control input using encrypted sampled data without decrypting them.
As the controller does not require a private key for decryption, encrypted control systems can not only protect the privacy of the plant data from the controller, but also enhance the cyber-security. 
The idea of encrypted control system was first proposed 
 based on public-key RSA \cite{RivSA78} and ElGamal \cite{Elg85} cryptosystems in \cite{KogF15}. Subsequently, an encrypted control system with Paillier cryptosystem \cite{Pai99} was  considered in \cite{FarSB17}, and a solution approach to quadratic optimization with Paillier cryptosystem was proposed in \cite{ShoGA16}.

This paper proposes encrypted control architectures using Paillier cryptosystem \cite{Pai99} \emph{combined with quantizers}.
As in \cite{KogF15}, the proposed architectures do not require the controller to have private keys to compute the control inputs.
The main contribution of this paper is to propose the augmentation of quantizers whose sensitivity changes while the system evolves.
The quantizers are applied to real-valued sampled data and map to integers in $[-q_{\text{sat}},q_{\text{sat}}]$ for a fixed saturation value $q_{\text{sat}}$.
Thus, the plaintext space (key length) may be kept small by choosing a small $q_{\text{sat}}$, which \emph{allows us to balance between cipher strength and control performance (sampling time)}. This is essential for control systems that require real-time computation of control inputs.
Moreover, the use of quantizers eliminates the analysis for the fixed-arithmetic  \cite{FarSB17}  to guarantee stability, and allows the linear system to 
 achieve asymptotic stability.  
Other contributions include extensions to event-triggered control and nonlinear control systems.
In particular, this is the first study to consider the construction of an encrypted nonlinear control system.

The rest of the paper is organized as follows. 
Section \ref{sec:prep} provides the mathematical preliminaries and operation rules of encrypted data (ciphertext).
An encrypted linear state-feedback control is presented in Section \ref{sec:sf}, which is extended to event-triggered control in Section \ref{sec:et} and nonlinear control in Section \ref{sec:nl}.
Finally, the paper is concluded in Section \ref{sec:con}.


\section{Preparations} \label{sec:prep}

\subsection{Notation}
The sets of real numbers and integers are denoted by $\mathbb{R}$ and  $\mathbb{Z}$, respectively.  
The set of vectors of length $n$ is denoted by $\mathbb{R}^n$ and the set of matrices of size $n$ by $m$ is denoted by $ \mathbb{R}^{n\times m}$.
The greatest common divisor and the least common multiple of $a, b \in \mathbb{Z}\setminus \{0\}$ are denoted by $\gcd(a,b)$ and $\operatorname {lcm}(a, b)$, respectively.
We define the sets of integers $\mathbb{Z}_n := \{z\in \mathbb{Z}: 0\leq z < n\}$ and $\mathbb{Z}_n^*: = \{z\in \mathbb{Z}_n:  \gcd(z,n) = 1\}$. 
For a vector $v \in \mathbb{R}^n$, the $i$th element of $v$ is denoted by $v_{i}$, and the Euclidean norm is denoted by   $\|v\|$. 
For a matrix $M \in \mathbb{R}^{n\times n}$, the $i,j$th element of $M$ is denoted by $m_{ij}$,  and  the induced 2-norm and the Frobenius norm are denoted by $\|M\|$ and  $\|M\|_F$, respectively.
 The maximum and minimum eigenvalues of a symmetric matrix $M=M^T$ are denoted by $\lambda_{\max}(M)$ and $\lambda_{\min}(M)$, respectively.
The floor function is denoted by $\lfloor x \rfloor:= \max\{k \in \mathbb{Z}: k < x\}$.

\subsection{Quantizer}\label{subsec:q}
Paillier cryptosystem operates over the message  of nonnegative integers (plaintext). 
However, the control theory usually deals with the data of real numbers.
In order to map the data to nonnegative integers, we use quantizers.

For a positive integer $q_{\text{sat}}$ and a positive real number $\Delta$, 
a quantizer $q: \mathbb{R} \rightarrow \mathbb{Z}$ is given by \cite{BroL00}
\begin{align}
\!\!\!q_{\Delta}(x) \! \!:=\! \!
\begin{cases}
q_{\text{sat}} &\! \! \! \! \!\!   \text{ if } x > (q_{\text{sat}}+1/2) \Delta,\\
-q_{\text{sat}} &\! \! \! \! \! \!  \text{ if } x \leq - (q_{\text{sat}}+1/2) \Delta,\\
\! \left\lfloor \! \dfrac{x}{\Delta}\!  +\! \dfrac{1}{2}\!  \right\rfloor
&\! \! \! \! \! \!  \text{ if} -\! (q_{\text{sat}}\! +\! 1/2) \Delta\! \!<\!\! x\! \! \leq (q_{\text{sat}}\! +\! 1/2) \Delta,
\end{cases}
\end{align}
where $\Delta$ is the sensitivity of the quantizer and $q_{\text{sat}}$ is the saturation value of the quantizer.

For the shorthand notation, we define
\begin{align}
x_q := q_{\Delta}(x), \ \bar{x}:=x_q\Delta, \ \tilde{x} := x-\bar{x}, \label{eq:xnotation} 
\end{align}
then
\begin{align}
 \bar{x} -\Delta/2 \leq x<  \bar{x}+  \Delta/2, \ |\tilde{x}| \leq \Delta/2. \label{eq:xrange} 
\end{align}
With an abuse of notation, we write $v_q:=q_{\Delta}(v)\in \mathbb{R}^n$ and $M_q:=q_{\Delta}(M)\in \mathbb{R}^{n\times m}$ to denote element-wise quantization with $\Delta$ for a vector $v$ and a matrix $M$, and define $ \bar{v}$, $\tilde{v}$, $\bar{M}$ and $\tilde{M}$ similarly to \eqref{eq:xnotation}.  
Then, 
\begin{align}
\|\tilde{v}\| \leq \Delta \sqrt{n}/2, \  \|\tilde{M} \| \leq \|\tilde{M} \|_{\text{F}} \leq \Delta \sqrt{n m}/2. \label{eq:q_err}
\end{align}

\subsection{Paillier cryptosystem}
An overview of Paillier cryptosystem \cite{Pai99} is given below.
\subsubsection{Encryption scheme}
\begin{itemize}
\item Key generation: 
\begin{itemize}
\item Choose two large prime numbers $p$ and $q$ randomly and independently of each other such that $ \gcd(pq,(p-1)(q-1))=1$
\item Generate public key: $N= pq$, and $g \in \mathbb{Z}_{N^2}^*$ such that the order of $g$ is a multiple of $N$
\item Generate private key: $ \lambda =\operatorname {lcm} (p-1,q-1)$
 \end{itemize}
\item Encryption (P-encryptor): Given a message $m\in  \mathbb{Z}_N$, 
\begin{itemize}
\item Compute ciphertext:  $c=g^{m}\cdot r^{N}{\bmod {N}}^{2}$ with a random integer $r \in \mathbb{Z}_N^*$
\end{itemize}
\item Decryption (P-decryptor): Given a  ciphertext $c<N^2$,
\begin{itemize}
\item Compute the message: $m=L(c^{\lambda }{\bmod  N}^{2})/L(g^{\lambda }{\bmod  N}^{2}) {\bmod N}$, where $L(x)=(x-1)/N$
\end{itemize}
\end{itemize}

We denote the Paillier encryption of the message $m$ by $\mathcal{E}_P(m)$, and the decryption of ciphertext $c$ by $\mathcal{D}_P(c)$.

\subsubsection{Encryption properties}
Paillier cryptosystem allows us to add two encrypted values with the addition operator $\oplus$ and to multiply by a plaintext with the multiplication operator $\otimes$. Namely, 
for $m, m_i\in \mathbb{Z}_N$,
it holds that
\begin{align}
\begin{aligned}
&\mathcal{D}_P\!\left( \!\mathcal{E}_{P}(m_1) \oplus \mathcal{E}_{P}(m_2) \bmod N^2 \right) \!=\! m _1+m_2 \bmod N,\\
&\mathcal{D}_P\left(a \otimes \mathcal{E}_{P}(m)\bmod N^2 \right) = a m \bmod N, 
\end{aligned}
\end{align}
therefore,
\begin{align}
&\mathcal{D}_P\left(\bigoplus_i a_i \otimes \mathcal{E}_{P}(m_i) \bmod N^2 \right) \!= \!\sum_i a_i m_i \bmod N.
\end{align}

\subsection{Multiplicative blinding using a random number (r-encryptor/r-decryptor)}
To perform some computations that can not be performed on Paillier encrypted values $\mathcal{E}_{P}(m)$, we must offload such computations. 
To avoid direct access,  as is often done, we ``obfuscate''  the message $m$ by multiplying a random number $r \in \mathbb{Z}_{r_{\max}}$ for some $r_{\max} \in \mathbb{Z}/\{0\}$ to obtain $\mathcal{E}_r(m) := rm$ (multiplicative blinding \cite{Ker11}). The inverse operation is denoted by $\mathcal{D}_r(c) := c/r$. 

\subsection{Matrix-vector multiplication}
The element-wise encryptions for a vector $v$ and a matrix $M$ are denoted by $\mathcal{E}_P(v)$ and  $\mathcal{E}_P(M)$, respectively, and the corresponding element-wise decryptions are denoted by $\mathcal{D}_P(v)$ and  $\mathcal{D}_P(M)$, respectively. 
The element-wise multiplicative blinding for a vector $v$ and a matrix $M$ are denoted by $\mathcal{E}_r(v):=rv$ and  $\mathcal{E}_r(M):=rM$, respectively, and the corresponding element-wise decryptions are denoted by $\mathcal{D}_r(v):=(1/r)v$ and  $\mathcal{D}_r(M):=(1/r)M$, respectively. 

With these notation, for $b_i\in \mathbb{Z}_N$, observe that
\begin{align}
\begin{aligned}
&\mathcal{D}_P\left(\mathcal{E}_{r}(a) \otimes \mathcal{E}_{P}(b) \bmod N^2 \right) =\mathcal{E}_{r}(a)b  \bmod N,\\
&\mathcal{D}_P\!\left(\!\! \bigoplus_i \mathcal{E}_{r}(a_i) \otimes \mathcal{E}_{P}(b_i)\bmod N^2 \!\! \right) \!=\!
 \sum_i  \mathcal{E}_{r}(a_i) \!\otimes\! b_i   \bmod N.
\end{aligned}
\end{align}
Therefore, for $\mathcal{E}_r(A)$ with a matrix $A \in \mathbb{Z}^{n\times n}$ and $\mathcal{E}_P(b)$ with a vector $b\in \mathbb{Z}^n$  such that $b_i\in\mathbb{Z}_N$, it holds that 
\begin{align}
\mathcal{D}_P\!\left(\!\left(\!\mathcal{E}_r(A) \!\otimes\!\mathcal{E}_P(b)\! \right)_{i} \bmod N^2 \!\right)\! =\! \sum_j\! \mathcal{E}_r(a_{ij}) b_j  \bmod N,
\end{align}
thus
\begin{align}\begin{aligned}
\mathcal{D}_P\left(\mathcal{E}_r(A) \otimes\mathcal{E}_P(b)
\bmod N^2 \right) &=
\mathcal{E}_r(Ab) \bmod N.\label{eq:multiplication}
\end{aligned}\end{align}
If $r\in \mathbb{Z}_N$ and each element of $rA b$ is in $\mathbb{Z}_N$, we have
\begin{align}
&\mathcal{D}_r\left(\mathcal{D}_P(\mathcal{E}_P\left(\mathcal{E}_r(A) b\mod N\right)\bmod N^2)  \right) 
= A b. \label{eq:multiplication_d}
\end{align}

\subsection{Remarks between quantizer and encryptor}\label{sec:qande}
After mapping a real number to an integer using the quantizer, we need to convert this integer to an element in $\mathbb{Z}_N$.
As Paillier arithmetic uses modulo $N$, we may take the convention that a number $x < N/3$ is positive, and that a number $x > 2N/3$ is negative. 
The range $N/3 < x < 2N/3$ allows for overflow detection \cite{PytP16}. 
We include this mapping from an integer to a nonnegative integer in the steps of encryption, i.e., we use $\mathcal{E}_P(x)$ to denote the Paillier encryption of $x \bmod N$. 
Similarly, we use $\mathcal{D}_P(x)$ to denote the Paillier decryption of $x$ followed by subtraction of $N$ if the Paillier decryption yields the value greater than $N/2$.

\section{Encrypted State-feedback Control}\label{sec:sf}
In this section, we present an encrypted linear state-feedback control system that achieves asymptotic stability.

Consider the discrete-time linear system
\begin{align}
\begin{aligned}
x[t+1] &= Ax[t]+Bu[t], \ \ t = 0,1,2,\cdots, \label{eq:sf}
\end{aligned}
\end{align}
where $x[t]\in \mathbb{R}^{n_x}$ is the system state, and $u[t]\in  \mathbb{R}^{n_u}$ is the control input. Suppose that $(A,B)$ is controllable, and  $K\in \mathbb{R}^{n_u \times n_x}$ is found  such that $A-BK$ is Schur, i.e., all the eigenvalues of $A-BK$ are inside the unit circle in the complex plane. 
Then, the system \eqref{eq:sf} is stabilized by the state-feedback control 
\begin{align}
u[t] &=- Kx[t]. \label{eq:sf_u}
\end{align}

\textbf{Problem: Design an encrypted control system for \eqref{eq:sf}-\eqref{eq:sf_u} that achieves the asymptotic stability while  protecting the privacy of  the sampled data $x[t]$ and the controller gain $K$ from the controller. }

In the following subsections, the overall control design is presented, followed by the analysis and design of quantizers.
\subsection{Overview of encryption architecture}

\begin{figure}[h]
\centering
\includegraphics[width=0.4\textwidth,viewport =210 135 800 495,clip]{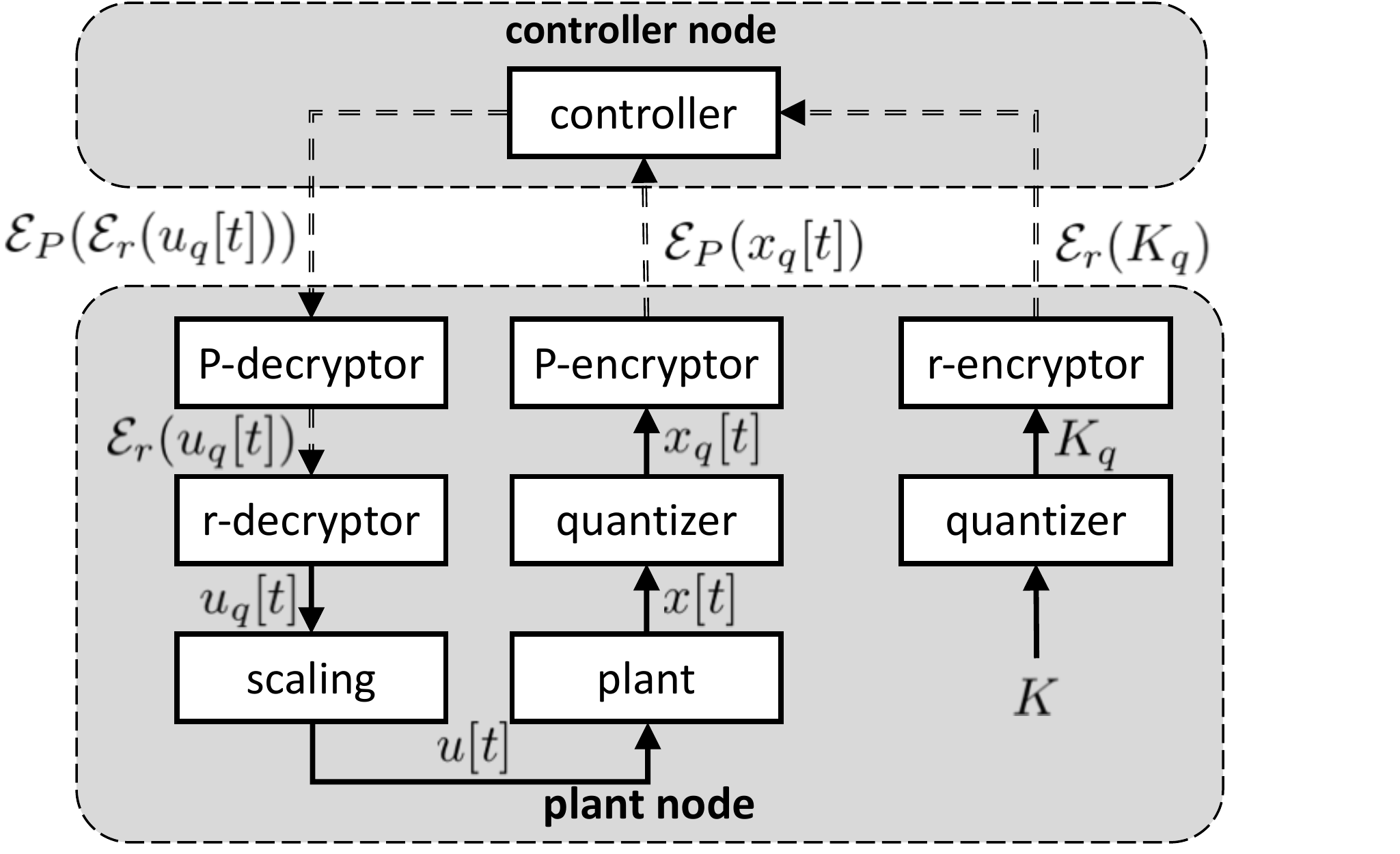} 
\caption{Proposed encrypted control architecture. Dashed lines indicate the flow of encrypted data and solid lines indicate the flow of plaintext data.  \label{fig:fig1}}
\end{figure}

The proposed architecture consists of the plant node and controller node between which no information from which the values of $x_q[t]$ and $K_q$ can be identified is exchanged (Figure \ref{fig:fig1}). In order to encrypt the sampled data $x[t]$ and the gain $K$, there are two quantizers in the system: one for $x[t]$ has the time-varying sensitivity $\Delta[t]$ and one for $K$ has the constant sensitivity $\Delta_g$, and both quantizers are designed not to saturate. 
 
The role of each node is summarized below (\textbf{P} and \textbf{C} denote the plant and controller nodes, respectively):
\begin{itemize}
\item[\textbf{P:}] Quantizes the gain $K$ to $K_q$, obfuscates $K_q$ to $\mathcal{E}_r(K_q)$, and then sends $\mathcal{E}_r(K_q)$ to the controller node. 
\item[\textbf{P:}] Quantizes the sampled state $x[t]$ of the plant to $x_q[t]$, encrypts $x_q[t]$ to $\mathcal{E}_P(x_q[t])$, and then sends $\mathcal{E}_P(x_q[t])$  to the controller node at every sampling time after some time $t_0$. Sends  $\mathcal{E}_P(0)$  to the controller node at every sampling time before $t_0$. (The time instance $t_0$ is determined in  \ref{sec:qins}.)
\item[\textbf{C:}]  Upon receiving the obfuscated/encrypted scaled data $\mathcal{E}_r(K_q)$ and $\mathcal{E}_P(x_q[t])$, computes the encrypted scaled control inputs $\mathcal{E}_P(\mathcal{E}_r(u_q[t]))$ for $u_q[t] = K_qx_q[t]$, and sends  $\mathcal{E}_P(\mathcal{E}_r(u_q[t]))$ to the plant node.
\item[\textbf{P:}]  Upon receiving the encrypted scaled control inputs $\mathcal{E}_P(\mathcal{E}_r(u_q[t]))$, decrypts $\mathcal{E}_P(\mathcal{E}_r(u_q[t]))$ to  $\mathcal{E}_r(u_q[t])$ then to $u_q[t]$.
\item[\textbf{P:}]  Scales $u_q[t]$ to obtain $u[t]=u_q[t]\Delta_g\Delta[t]$
using the sensitivities $\Delta_g$ and $\Delta[t]$ and applies it to the plant.
\end{itemize}

Note that the controller uses the encrypted data of the quantizer output (scaled approximations).
This means that the data sent and received among the three node is encryptions of integers between $-q_{\text{sat}}$ and $q_{\text{sat}}$.
Thus, there are no fractional bits, which renders multiplication easy \cite{FarSB17}.

This architecture preserves the privacy of the plant state $x[t]$ ($x_q[t]$) and the controller gain $K$ ($K_q$)  from the controller node because the controller does not know the private key and the value of $r$. Thus, if quantizers are designed such that the closed-loop achieves the asymptotic stability, then 
the overall encrypted control system achieves asymptotic stability while protecting the privacy of $x[t]$ and $K$ in the sense that the controller node can access only the encrypted data of $x[t]$ and $K$.

\begin{remark}$ \ $
\begin{itemize}
\item When generating the public key, $p$ and $q$ are chosen such that $N$ for the cryptosystem satisfies $N> 3(q_{\text{sat}}+1/2)(q_{\text{sat,g}}+1/2) n_xr_{\max}$, where $q_{\text{sat}}$ and $q_{\text{sat,g}}$ are the saturation values of the quantizers for the states and the controller gain, respectively, and determined in the next subsection.
This guarantees that the elements of $rK_qx_q \bmod N$ are uniquely determined for each $rK_qx_q$ and vice versa. 
Recalling the notation for $\mathcal{E}_{P}$ and $\mathcal{D}_{P}$ in Section \ref{sec:qande} and \eqref{eq:multiplication_d},   for
\begin{align*}
\mathcal{E}_{P}\left(\mathcal{E}_{r}(u_q)\right) =\mathcal{E}_{r}(K_q)\otimes \mathcal{E}_{P}\left(x_q\right)=\mathcal{E}_{P}\left(rK_qx_q\right),
\end{align*}
we have
\begin{align*}
\mathcal{D}_{r}\!\left(\mathcal{D}_{P}\left(\mathcal{E}_{P}\left(\mathcal{E}_{r}(u_q)\right)\right)\right)
\!=\! \dfrac{1}{r}\!\left(\mathcal{D}_{P}\left(\mathcal{E}_{P}\left(rK_qx_q\right)\right)\right)\!=\! K_qx_q.
\end{align*}
Similar assumptions on $N$ are posed on later sections.
\item The sensitivity of the quantizer for $K$ can be time-varying. For example, it can be synchronized with the sensitivity of the quantizer in the plant node as long as \eqref{eq:sen_K}  is satisfied.
\item The quantization and encryption of $K$ are required only once, but can be repeated using different random number at every sampling time.
\end{itemize}
\end{remark}

\begin{remark}$ \ $
Strictly speaking, the quantization and encryption of the state $x[t]$ is performed at sensor, 
which differs from where the quantization and encryption of the gain $K$ is performed. 
Similarly, decryptions and scaling to obtain the control input $u[t]$ is performed at actuator.
Those are combined in the plant node to indicate that they have common keys and quantizers.
\end{remark}
\subsection{Quantizer design}
To analyze the effect of quantization in the design of quantizers, consider the system  in Fig. \ref{fig:fig1} without encryptors and decryptors.
The quantized closed-loop system for \eqref{eq:sf}-\eqref{eq:sf_u} is 
\begin{align}
\begin{aligned}
x[t+1] =& Ax[t]+Bu[t] = Ax[t]-B\bar{K}\bar{x}[t].
\label{eq:sf_q}
\end{aligned}
\end{align}

\subsubsection{Quantizer for the gain}
Let us first determine the sensitivity $\Delta_g$ of the quantizer for the gain $K$.

As $A-BK$ is Schur, for any given $Q =Q^T>0$, there exists $P=P^T>0$ such that 
\begin{align}
(A-BK)^TP(A-BK)-P+Q=0. \label{eq:q}
\end{align}
It is guaranteed that $A-B\bar{K}$ is Schur if
\begin{align}\begin{aligned}
&(A-B\bar{K})^TP(A-B\bar{K})-P\\
=&(A-BK+B\tilde{K})^TP(A-BK+B\tilde{K})-P\\
\leq& -Q+(A-BK)^TPB\tilde{K}+\tilde{K}^TB^TP(A-BK)\\
&\qquad +\tilde{K}^TB^TPB\tilde{K}\leq -\varepsilon I, \ \varepsilon>0.
\end{aligned}\end{align}

Therefore, after some computations using \eqref{eq:q_err}, the sensitivity $\Delta_g$ is chosen to satisfy
\begin{align}\begin{aligned}
&\Delta_g \leq \varepsilon'+  \dfrac{2}{\sqrt{n_xn_u}\|B^TPB\|} \left(- \|(A-BK)^TPB\| \right.\\ & \left. \quad +\sqrt{ \|(A-BK)^TPB\|^2+\lambda_{\min}(Q)\|B^TPB\| }\right), \label{eq:sen_K} 
\end{aligned}\end{align}
for any $Q =Q^T>0$ of the designer's choice and the corresponding $P$ satisfying \eqref{eq:q} with  $\varepsilon'>0$.

Once the sensitivity is determined, the saturation value $q_{\text{sat},g}$ is selected such that 
the elements of $K$ are not truncated, i.e., 
\begin{align}
&\max_i \max_j |K_{ij}| \leq (q_{\text{sat},g}-1/2) \Delta_g. \label{eq:sat_K} 
\end{align}

To simplify the notation, once the quantizer is determined and the gain is quantized to $\bar{K}=K_q\Delta_g$, we choose $\bar{Q}=\bar{Q}^T>0$ and find $\bar{P}=\bar{P}^T>0$ that solves
\begin{align}
(A-B\bar{K})^T\bar{P}(A-B\bar{K})-\bar{P}+\bar{Q}=0.
\end{align}
We use this $\bar{P}$ and $\bar{Q}$ to design the quantizer for the states.

\subsubsection{Quantizer for the states} \label{sec:qins}
The design of quantizer for the states follows the approach proposed in \cite{BroL00}. Due to space limitation, we only present the summary of the quantizer and interested readers are referred to the cited reference. 

The employed quantizer uses a constant saturation level $q_{\text{sat}}$ and a time-varying sensitivity given by
\begin{align}
\begin{aligned}
\Delta[t] =
\begin{cases}
 \|A\|^{2t}, & 0\leq t <t_0\\
\Delta_i :=  \Omega^i \Delta_0, & t_0 \leq t_i \leq t < t_{i+1}, \label{eq:tv_q}
\end{cases}
\end{aligned}
\end{align}
where $\Delta_0 = \|A\|^{2t_0}$, $\Omega$ is a scaling factor and $t_i$ are the  time instances of sensitivity updates. 
Thus, the sensitivity decreases by the factor of $\Omega$ at every time of updates.

The scaling factor is given by 
\begin{align}
\Omega := \Omega' \sqrt{\dfrac{\lambda_{\max}(\bar{P})}{\lambda_{\min}(\bar{P})}}\left(q_{\text{sat}}-\dfrac{1}{2}\right)^{-1}, \label{eq:sat_omega}
\end{align}
where
\begin{align}
&\begin{aligned}
\Omega' :=\left(\Theta\sqrt{n_x}+ \varepsilon\right) \sqrt{\dfrac{\lambda_{\max}(\bar{P})}{\lambda_{\min}(\bar{P})}} +\sqrt{n_x},\\
\end{aligned}\\
&\begin{aligned}
&\Theta := \dfrac{1}{2\lambda_{\min}(\bar{Q})}\left(\|(A-B\bar{K})^T\bar{P}B\bar{K}\| \right.\\
& \left.\! \!+\!\sqrt{\!\|(A\!-\!B\bar{K})^T\bar{P}B\bar{K}\|^2\!+\!\lambda_{\min}(\bar{Q})\|\bar{K}^TB^T\bar{P}B\bar{K}\! \|}\!\right)\!, \label{eq:theta}
\end{aligned}\end{align}
and parameters $\varepsilon>0$ and $q_{\text{sat}}\geq 1$ are chosen such that $\Omega \in (0,1)$.

The time instances $t_i$ are given by 
\begin{align}
\begin{aligned}
\ t_{0} :=  &\min \left\{ t \geq 1: \left\|q_{\Delta[t]}(x[t]) \right\| \right.\\
&\qquad \left. \leq \left(q_{\text{sat}}-\dfrac{1}{2}\right) \sqrt{\dfrac{\lambda_{\max}(\bar{P})}{\lambda_{\min}(\bar{P})}}  -\dfrac{\sqrt{n_x}}{2}\right\},\\
t_{i+1} :=  &\min \left\{ t \geq t_{i}+ 1: \left\|q_{\Delta_i}(x[t]) \right\|\leq \!\Omega' \!-\dfrac{\sqrt{n_x}}{2}\right\}\!. 
\end{aligned}\label{eq:updates}\end{align}

By construction, it holds that
\begin{align}\begin{aligned}
&\|x[t_0] \| \leq \Delta_0 \left(q_{\text{sat}}-\dfrac{1}{2}\right) \sqrt{\dfrac{\lambda_{\min}(\bar{P})}{\lambda_{\max}(\bar{P})}},\\
&\|x[t_i] \| \leq \Delta_i \Omega' , \ i = 1, 2,\cdots.
\end{aligned} \end{align} 
and this quantizer guarantees that 
\begin{align}
x[t] \! \in\! R_{i+1}\! :=\! \left\{\!x\!: \!x^T\bar{P}x \leq {\lambda}_{\min}(\bar{P}) \Delta_{i}^2\left(\!q_{\text{sat}}-1/2\right)^2\!\right\} \label{eq:x_i_set}
\end{align} 
for $t\in [t_{i},t_{i+1})$.

These quantizers lead to asymptotic stability of the closed-loop system because
the rule for sensitivity updates \eqref{eq:tv_q} implies $\Delta[t] \rightarrow 0$ as $t\rightarrow \infty$, and 
\eqref{eq:x_i_set} implies that $x[t]$ approaches to $0$ as $\Delta[t] \rightarrow 0$.


\section{Extension to Event-triggered Control}\label{sec:et}

This section presents how to augment an event-triggered control scheme to the encrypted control law developed in Section \ref{sec:sf} to save communications and actuator updates. 
Event-triggered control takes samples of the plant state at every time instance and updates the control input only when specified conditions are satisfied \cite{HeeJT12}.
 
\textbf{Problem: Design an event-triggered encrypted control system for \eqref{eq:sf}-\eqref{eq:sf_u} that achieves the asymptotic stability while  protecting the privacy of the sampled data $x[t]$ and the controller gain $K$ from the controller. }

We propose to augment an event-trigger architecture to the plant node.
More specifically, we implement the event-trigger mechanism between the plant and the quantizer in the plant node.
This way, the sampled data is quantized, encrypted and sent to the controller only when an event-trigger condition is met.
In the following, the event-trigger condition is designed.

The event-triggered control system is given by
\begin{align}\begin{aligned}
x[t+1] &= Ax[t]+Bu[t], \\
  u[t] &= -\bar{K}\bar{x}[t^{(i)}], \ t^{(i)} \leq t < t^{(i+1)}.
\end{aligned}\end{align}
where $t^{(i)}$ for $i = 1,2,\cdots$, are the time instances of the control updates, and
 $\bar{x}[t^{(i)}]: = x_q[t^{(i)}]\Delta[t^{(i)}]$.

Using a Lyapunov function $V[t]=x^T[t]\bar{P}x[t]$,
\begin{align*}
&V[t+1]-V[t] 
\leq - \lambda_{\min}(\bar{Q}) \|x\|^2\\&+2\|(A-B\bar{K})^T\bar{P}B\bar{K}\|\|x\|\|e\|
 +\|\bar{K}^TB^T\bar{P}B\bar{K}\|\|e\|^2,
\end{align*}
where
$e[t] = x[t]-\bar{x}[t^{(i)}]$.
This Lyapunov function is negative outside the ball $\{x: \|x\|\leq 2 \Theta \|e\|\}$, where $\Theta$ is in \eqref{eq:theta}.

Thus, an event-trigger condition can be set as
\begin{align}
t^{(i+1)} = \min\left\{t \geq t^{(i)}: \|x[t]\|\leq 2\Theta \|e[t]\| \right\}. \label{eq:et}
\end{align}

When the event-trigger condition is satisfied, check if the quantizer needs to be updated or not, and update if necessary using \eqref{eq:updates}. The rest of the analysis is the same as in Section \ref{sec:sf} and  \cite{BroL00}. This is because the aforementioned event-trigger condition \eqref{eq:et} guarantees the decrease of the Lyapunov function, based on which the analysis is developed. 

This is a straightforward extension of well-known results, because the plant node knows both $x[t^{(i)}]$ and $x[t]$.

\begin{remark}
It is also possible to augment an event-trigger architecture to the controller node rather than the plant node.
However, in order to do this, it is needed to add another node that communicates with the controller and checks the satisfaction of the event-triggered condition. 
\end{remark}

\section{Extension to Nonlinear Systems} \label{sec:nl}
This section extends the approach in Section \ref{sec:sf} to a simple nonlinear system using feedback linearization \cite{Kha01}.

Consider the scalar nonlinear system
\begin{align}
x[t+1] =ax[t] + b (u[t]-\alpha (x[t])), \ t=0,1,2,\cdots, \label{eq:sf_n}
\end{align}
where $x[t] \in \mathbb{R}$ is the system state, and $u[t]\in  \mathbb{R}$ is the control input. 
Assume that $ab \neq 0$. 

The feedback linearization uses the control input $u[t]= \alpha (x[t])-v[t]$, yielding 
\begin{align}
\begin{aligned}
x[t+1] 
&= ax[t]-bv[t]. \label{eq:sf_nv}
\end{aligned}
\end{align}
If  $k\in \mathbb{R}$ such that $|a-bk|< 1$ is selected,  
then, the system \eqref{eq:sf_nv} is stabilized by 
$v[t] = kx[t]$,
and  \eqref{eq:sf_n} is stabilized by 
\begin{align}
u[t] &=\alpha (x[t])- kx[t].\label{eq:sf_nu}
\end{align}

\textbf{Problem: Design an encrypted control system for \eqref{eq:sf_n} using \eqref{eq:sf_nu} that achieves the practical stability while  protecting the privacy of  the sampled data $x[t]$ on a bounded set $\mathcal{X}:=[x_{\min}, x_{\max}]$ from the controller. }

The system is said to be practically stable if $|x[0]|< c_1$, then $|x[t]|< c_2$ for $t\geq \bar{t}$ for some $\bar{t}>0$ for given $c_1$ and $c_2$ such that $0<c_1<c_2$ \cite{LakLM90}.
The reason for requiring practical stability instead of asymptotic stability will become clear in the rest of this section.

\subsection{Function approximation}
In order to compute the control input using encrypted data for $x[t]$, we first approximate the nonlinear function $\alpha (x[t])$ using the quantized values.
From Weierstrass approximation theorem \cite{Est02}, for any $\varepsilon_1'>0$ there exist $p$ and $c_j$ such that $\alpha_p(x):= \sum_{j=0}^{p} {c}_j x^j$ satisfies
\begin{align}
| \alpha_p(x) -\alpha (x)| \leq \varepsilon_1', \ \forall x \in \mathcal{X}. \label{eq:e1}
\end{align}
With a quantizer of sensitivity $\Delta$, define
\begin{align}\begin{aligned}
&\bar{\alpha}_p (\bar{\mathbf{x}}) := \sum_{j=0}^{p} \bar{c}_j\bar{x}^{(j)}={\mathbf{c}}_q^T{\mathbf{x}}_q\Delta^2,
\end{aligned}\end{align}
where $c_{j,q} \!= \!q_{\Delta}(c_j)$, $x^j_q \!=\! q_{\Delta}(x^j)$ as usual, and 
\begin{align}\begin{aligned}
\bar{c}_j \!&:=\! c_{j,q} \Delta,\  \bar{x}^{(j)} \!:=\! x^j_q \Delta, \\
{\mathbf{c}} \!&:=\! \left[\! \!  \begin{array}{cccc} \!  c_0 \! \! &\! c_1\!\!  &\!  \cdots \! \! &c_{p} \! \end{array}\! \! \right]^T\!\!\! \!\!, \ \mathbf{x} \!:=\! \left[\! \!  \begin{array}{cccc} \!  1 \! \!&\! {x}\!  &\!  \cdots \! \! &{x}^{p} \! \end{array}\! \! \right]^T\!\! \! \!,\\
 {\mathbf{c}}_q\! &:=q_{\Delta}(\mathbf{c}), \ {\mathbf{x}}_q \!:=q_{\Delta}(\mathbf{x}), \ \bar{\mathbf{x}} = {\mathbf{x}}_q\Delta.
\end{aligned}\end{align}
Then with some constant $\varepsilon_2$, it holds that
\begin{align}\begin{aligned}
&| \bar{\alpha}_p (\bar{\mathbf{x}})-\alpha_p(x)| \leq \sum_{j=0}^{p} |\bar{c}_j-{c}_j| |\bar{x}^{(j)}|\\& \quad
+ |{c}_j -\bar{c}_j|| \bar{x}^{(j)}-x^j|+| \bar{c}_j||\bar{x}^{(j)}-x^j| \leq \varepsilon_2 \Delta/2. \label{eq:e2}
\end{aligned}\end{align}
With $\varepsilon_1 = 2\varepsilon_1'/\Delta$,  \eqref{eq:e1} and \eqref{eq:e2} imply that
\begin{align}
| \bar{\alpha}_p (\bar{\mathbf{x}}) -\alpha (x)| \leq M \Delta/2, \ M:= \varepsilon_1+\varepsilon_2.
\end{align}

\subsection{Overview of encryption architecture}
\begin{figure}[h]
\centering
\includegraphics[width=0.4\textwidth,viewport =210 139 800 495,clip]{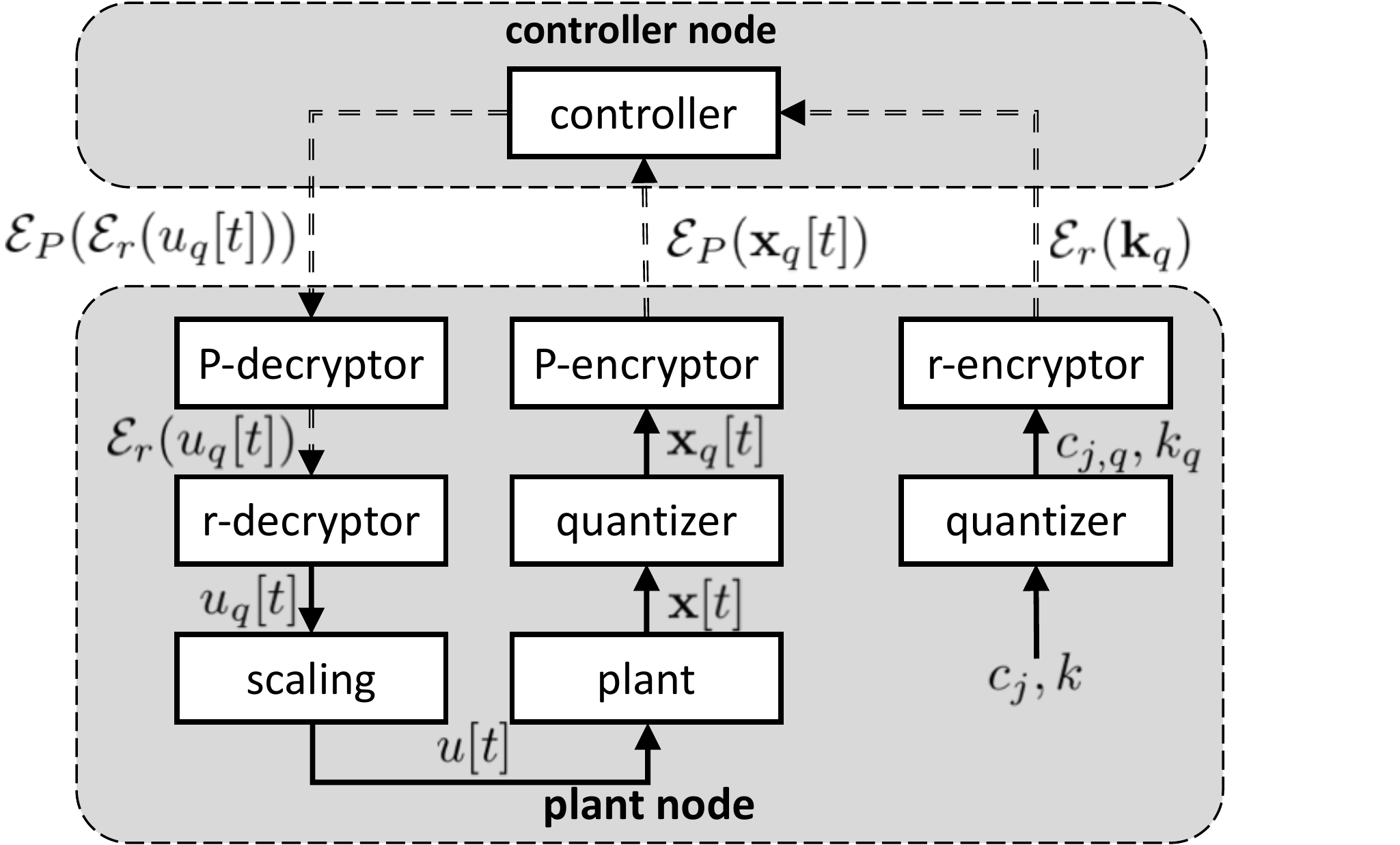} 
\caption[something]{Proposed encrypted control architecture. Dashed lines indicate the flow of encrypted data and solid lines indicate the flow of plaintext data.  \label{fig:fig2} } \vspace{-0.2in}
\end{figure}

As before, the proposed architecture consists of two nodes between which only encrypted data is exchanged (Figure \ref{fig:fig2}). 
However, two quantizers maintain the same sensitivity $\Delta[t]$. 

The role of each node is summarized below (\textbf{P} and \textbf{C} denote the  plant and controller nodes, respectively):
\begin{itemize}
\item[\textbf{P:}] Quantizes the gain $k$ and the coefficients $c_j$ to $k_q$ and $c_{j,q}$, respectively, and
constructs a vector $\mathbf{k_q} ={\mathbf{c}}_q-{k}_q\mathbf{e}_2$, where  $\mathbf{e}_2$ is the second column of the identity matrix of size $p+1$.
Then,  obfuscates $\mathbf{k_q}$ to $\mathcal{E}_r(\mathbf{k_q})$, and then sends  $\mathcal{E}_r(\mathbf{k_q})$ to the controller node (at every time the sensitivity changes). \item[\textbf{P:}] Quantizes the polynomial basis of the sampled state $\mathbf{x}[t]$ to ${\mathbf{x}}_q[t]$, and
encrypts ${\mathbf{x}}_q[t]$ to  $\mathcal{E}_P({\mathbf{x}}_q[t])$, and then sends  $\mathcal{E}_P({\mathbf{x}}_q[t])$  to the controller node (at every sampling time). 
\item[\textbf{C:}]  Upon receiving the obfuscated/encrypted  data, computes the scaled encrypted  control inputs $\mathcal{E}_P(\mathcal{E}_r(u_q[t]))$ for $u_q[t] = \mathbf{k_q}^T{\mathbf{x}}_q$. 
\item[\textbf{P:}]  Upon receiving the encrypted scaled control inputs $\mathcal{E}_P(\mathcal{E}_r(u_q[t]))$, decrypts $\mathcal{E}_P(\mathcal{E}_r(u_q[t]))$ to  $\mathcal{E}_r(u_q[t])$ and then $u_q[t]$.
\item[\textbf{P:}]  Scales $u_q[t]$ to obtain the control input  $u[t]=u_q[t]\Delta^2[t]$ and applies it to the plant.
\end{itemize}

\subsection{Quantizer analysis and design}
As before, we analyze the effect of quantization  in the design of quantizers by considering the system in Fig. \ref{fig:fig2} without encryptors and decryptors. 

With the quantized control input
\begin{align}
u[t] &=\bar{\alpha}_p (\bar{\mathbf{x}}[t])- \bar{k}\bar{x}[t],
\end{align}
the quantized closed-loop system is
\begin{align}
\begin{aligned}
x[t+1] 
&= ax[t]-b \bar{k}\bar{x}[t] +b(\alpha (x[t]) -\bar{\alpha}_p (\bar{\mathbf{x}}[t])).
 \label{eq:sf_cl}
\end{aligned}
\end{align}
With a Lyapunov function $V =(x[t])^2$, \eqref{eq:sf_cl} implies that
\begin{align}
\begin{aligned}
&V[t+1]-V[t]  \\
&= (ax[t]-b \bar{k}\bar{x}[t] +b(\alpha (x[t]) -\alpha (\bar{x}[t])))^2-x^2[t]\\
&\leq((a-b\bar{k})^2-1)x^2[t]+2(a-b\bar{k})y[t]x[t] +y^2[t],
\end{aligned} \label{eq:sf_ly}
\end{align}
where 
$
y[t] = b(\bar{k}+M)\Delta[t]/2$.

The expression \eqref{eq:sf_ly} is negative outside the ball $\{x: |x| \leq \Theta\Delta\}$, where
$
\Theta = (b(\bar{k}+M))/(1-(a-b\bar{k}))$.

This time, consider using a modified version of the quantizer in Section  \ref{sec:sf}, i.e., 
\begin{align}
\begin{aligned}
\Delta[t] =\Delta_i :=  \Omega^i \Delta_0, \  t_0 =0\leq t_i \leq t < t_{i+1}, \label{eq:tv_q_nl}
\end{aligned}
\end{align}
where $\Omega$ is a scaling factor and $t_i$ are the  time instances of sensitivity updates.

Unlike Section \ref{sec:sf}, the quantizer is initialized with the sensitivity $\Delta_0$ and the saturation level $q_{\text{sat}}$ such that satisfy
\begin{align}
|x[0]|\!\leq \!\Delta_0\!\left(\!q_{\text{sat}}-\dfrac{1}{2}\right)\!\!, \
\Omega \!=\! (\Theta+ \varepsilon +1)\! \left(\!q_{\text{sat}}-\dfrac{1}{2}\right)^{\!\!-1}\!\!\!\!\!\!\!< \!1, 
\end{align}
with some $\varepsilon>0$.
Then, we have $\Theta< q_{\text{sat}}-1/2$.

In order to avoid truncating $k$ and $c_j$, make sure that $q_{\text{sat}}$ is large enough satisfying
\begin{align}
|k| \leq (q_{\text{sat}}-1/2)\Delta_0, \ |c_j| \leq (q_{\text{sat}}-1/2)\Delta_0, \ \forall j.
\end{align}

Also in order to guarantee $|a-b\bar{k}|< 1$,  make sure that
\begin{align}\begin{aligned}
&\Delta_0 \leq \varepsilon'+  \dfrac{2\left(1- |a-bk|  \right)}{b} , \   \varepsilon'>0.
\end{aligned}\end{align}

Choosing the time instances of sensitivity updates 
\begin{align}
t_{i+1} \!= \! \min \left\{ t \geq t_{i}+ 1:\! \left\|q_{\Delta_i}(x[t]) \right\|\leq \Theta+ \varepsilon +1/2\right\}\!\!, \label{eq:updates_nl}
\end{align}
it holds that
\begin{align}
&|x[t_i] | \leq \Delta_i\left(\Theta+ \varepsilon +1\right) , \ i = 1, 2,\cdots.
\end{align} 
The existence of $t_i$ is guaranteed using the similar analysis in  \cite{BroL00} while $k$ and $c_j$ are not truncated.

This quantizer guarantees that 
\begin{align}
x[t] \in R_{i+1} := \left\{x: |x| \leq \Delta_{i} \left|q_{\text{sat}}-1/2\right| \right\} \label{eq:x_i_set_nl}
\end{align} 
for $t\in [t_{i},t_{i+1})$ as long as $k$ and $c_j$ are not truncated, i.e.,
\begin{align}
|k| \leq (q_{\text{sat}}-1/2)\Delta[t], \ |c_j| \leq (q_{\text{sat}}-1/2)\Delta[t], \ \forall j.
\end{align}

However, as $\Delta[t]$ approaches to zero, two problems occur:
\begin{itemize}
\item the quantized values of $k$ and $c_j$ will be truncated no matter how large $q_{\text{sat}}$ is chosen, and
\item the required upper bound $\varepsilon_1'$ for the function approximation \eqref{eq:e1} approaches to zero, which possibly leads to an infinitely large $p$.
\end{itemize}
Therefore, asymptotic stability cannot be guaranteed.
On the other hand, we may hold the sensitivity $\Delta[t]$ constant once it becomes sufficiently small to avoid the above two problems.
In other words, using the quantizer in the form of
\begin{align}
\!\!\!\!\Delta[t] \!=\!\!
\begin{cases}
\!\Delta_i \!:=\! \Omega^i \Delta_0,  t_0 \leq t_i \leq t < t_{i+1},  i = 1,\cdots, f \\
\!\Delta_f \!:= \! \Omega^f \Delta_0,   t_f \leq t, 
\!\!\end{cases}\!\!\!\label{eq:tv_q_nl}
\end{align}  
we can guarantee the practical stability of the system with 
\begin{align}
x[t] \in R_{f+1} := \left\{x: |x| \leq \Delta_{f} \left|q_{\text{sat}}-1/2 \right| \right\}, \ t \geq t_f,
\end{align} 
without incurring the problem.

We may also choose to use a time-invariant quantizer in the gain node to guarantee the practical stability, in which case, the region that $x[t]$ will stay depends on the sensitivity of the quantizer in the gain node.

\section{Conclusions}\label{sec:con}
In this paper, the control systems combined with quantizers and encryptors/decryptors are proposed and investigated.
It is shown that encrypted control systems can be constructed that achieve asymptotic stability for linear systems, and practical stability
for some nonlinear systems with the aid of function approximations using Weierstrass approximation theorem.
Since the combination with quantizers allows us to choose short key length, the processing time for encryption/decryption may be reduced for the sake of cipher strength. 


\addtolength{\textheight}{-15cm}   
\section*{ACKNOWLEDGMENT}
This research was supported by the grant from Okawa Foundation for Information and Telecommunications.

The author would like to thank Prof. Kiminao Kogiso at the University of Electro-Communications for his comments on the manuscript.

\bibliographystyle{IEEEtran}
\bibliography{IEEEabrv,myref}



\end{document}